\documentstyle[aps]{revtex}

\begin{document}

\headsep 2cm

\title{
Generic composition of boosts: an elementary derivation of the
Wigner rotation  \vskip1cm}

\author{\noindent\ Rafael Ferraro $^{1,2}\, $ \thanks{
email: ferraro@iafe.uba.ar} \ and Marc Thibeault  $^2\, $ \thanks{
email: marc@iafe.uba.ar} \\ \bigskip  \bigskip  $^1\, $ Instituto de
Astronom\'\i a y F\'\i sica del Espacio,\\ Casilla de Correo 67,2
Sucursal 28, 1428 Buenos Aires, Argentina \bigskip \\ $^2 \,$
Departamento de F\'\i sica, Facultad de Ciencias Exactas y
Naturales,\\ Universidad de Buenos Aires, Ciudad Universitaria,
Pabell\' on I,\\ 1428 Buenos Aires, Argentina}


\maketitle

\vskip2cm


\begin{abstract}
Because of its apparent complexity, the discussion of Wigner rotation is
usually reduced to the study of Thomas precession, which is too specific
a
case to allow a deep understanding of boost composition. However, by
using
simple arguments and linear algebra, the result for the Wigner rotation
is
obtaines straightforwardly, leading to a formula written in a manageable
form. The result is exemplified in the context of the aberration of
light.
\end{abstract}

\section{Introduction}

One of the most puzzling phenomenon in Special Relativity is the
composition of boosts. When one contemplates the form of an
arbitrary boost \cite{jackson}, it becomes clear that the
expression for the composition of two generic boosts will be very
complicated. As is known, the composition of boosts does not
result in a (different) boost but in a Lorentz transformation
involving rotation ({\it Wigner rotation }\cite{wigner}),Thomas
precession being the example normally worked out in the textbooks
\cite{jackson} ,\cite{Goldstein}, \cite{eisberg}, \cite{taylor}.
In this example, one is composing two boosts along mutually
perpendicular directions; for small velocities a second-order
approximation allows to get a result that is appropriate to
understand the precession of the spin of an electron inside an
atom.

Of course, the composition of two arbitrary boosts is also studied in
the
literature \cite{barut},\cite{ajp1},\cite{ajp2}, but generally the
treatments are too involved to capture the Wigner rotation easily.
Sometimes
the papers are aimed at the understanding of certain properties of the
Lorentz group, instead of looking for a straightforward way to get the
Wigner rotation, leaving in the reader the impression that this topic is
complicated, and cannot be comprehended without an involved analysis.
Moreover, the expressions are often difficult to use in practice, and
the
concepts are frequently hidden behind the abundance of mathematics. The
composition of boost and the Wigner rotation are therefore virtually
absent
from textbooks (save for the very specific case of Thomas precession).
One
is then left with the impression that the subject is subtle and
difficult.
Of course, this is true but not to the point of preventing its treatment
with simple mathematical tools.

In this paper the aim will be different. Our prime interest is in
the Wigner rotation; we choose the composition of boost as a
specific issue because some characteristics of boosts are
highlighted particularly well, the power of linear analysis is
demonstrated at its best, and, of course, because it is
interesting in itself. The mathematical tool that we will use is
simple linear algebra. After all, boosts are linear
transformations. However, the key point is that boosts are
symmetric linear transformations. This simple property will allow
us to effortlessly compute the Wigner rotation (see Eq.(
\ref{wigner}) below). Moreover, the understanding of the reason
that makes the boost symmetric will reveal some simple, basic
facts that are often passed over in textbook treatments. A second
goal of this paper is to present simple formulas to compute the
Wigner rotation. Their simplicity does not reside in their
explicit form; the final result will always be messy. However, we
want to give equations that are operationally simple in order that
the computation of the Wigner rotation should be a simple ``plug
and play'' procedure.

\section{Boost composition}

We will start by considering the composition of two boosts along
mutually
perpendicular directions. Before embarking upon calculation, one should
be
sure about what is looking for: one is wondering whether the composition
is
equivalent to a single boost or not. There are various ways of
understanding
this topic, depending to a large degree on the particular expertise and
taste of the reader. For the moment we will content ourselves with a
mathematical explanation. In Section III, we will clarify the meaning of
the
Wigner rotation by a physical exemple concerning the aberration of
light.

One could give an answer to the question by starting from the fact
that boosts are represented by symmetric matrices. On the one hand
one knows that a boost $B_x$ along the $x$ axis  is actually
represented by a symmetric matrix, and on the other hand one could
get a generic boost by performing an arbitrary spatial rotation:
$B_x\longrightarrow {\cal R}B_x{\cal = R}^{-1}$. Since the
rotations are orthogonal matrices, then a boost along an arbitrary
direction is also represented by a symmetric matrix $B={\cal
R}B_x{\cal R}^T$ $(B^T=B)$, whose form can be found in the
literature \cite{jackson}. This symmetry can also be regarded as a
reflection of the fact that boosts leave four independent
directions in spacetime invariant: namely, i) they do not modify
the light-cones; on the light-cone there are two independent
directions, belonging to light-rays travelling back and forth
along the boost direction, that remain invariant (see Appendix A);
ii) in addition, the spacelike directions that are perpendicular
to the boost direction are also left unchanged (a further two
independent directions). Then, boosts have four independent real
eigen(four)-vectors, and their representative matrices must be
symmetric (i.e., diagonalizable). In contrast, a (spatial)
rotation changes the directions belonging to the plane where it is
performed.

Since the product of matrices representing boosts is non-symmetric
(unless
both boosts are parallel), then one can answer that the composition of
two
boosts is not, in general, equivalent to a single boost. So we are
compelled
to analyze the result of the composition of two boosts as being
equivalent
to the composition of a boost and a rotation. Again the symmetry of
boosts
will allow us to identify the rotation in the result.

\subsection{Composition of mutually perpendicular boosts}

Let there be two boosts matrices along the $x$ and $y$ directions

\begin{equation}
B_{(x)}=\left(
\begin{array}{cccc}
\gamma _1 & -\gamma _1\beta _1 & 0 & 0 \\
-\gamma _1\beta _1 & \gamma _1 & 0 & 0 \\
0 & 0 & 1 & 0 \\
0 & 0 & 0 & 1
\end{array}
\right) ,  \label{boostx}
\end{equation}

\medskip\

\begin{equation}
B_{(y)}=\left(
\begin{array}{cccc}
\gamma _2 & 0 & -\gamma _2\beta _2 & 0 \\
0 & 1 & 0 & 0 \\
-\gamma _2\beta _2 & 0 & \gamma _2 & 0 \\
0 & 0 & 0 & 1
\end{array}
\right) .  \label{boosty}
\end{equation}

\bigskip

The product of these two matrices yields

\begin{equation}
B_{(y)}B_{(x)}=\left(
\begin{array}{cccc}
\gamma _2 & 0 & -\gamma _2\beta _2 & 0 \\
0 & 1 & 0 & 0 \\
-\gamma _2\beta _2 & 0 & \gamma _2 & 0 \\
0 & 0 & 0 & 1
\end{array}
\right) \left(
\begin{array}{cccc}
\gamma _1 & -\gamma _1\beta _1 & 0 & 0 \\
-\gamma _1\beta _1 & \gamma _1 & 0 & 0 \\
0 & 0 & 1 & 0 \\
0 & 0 & 0 & 1
\end{array}
\right) =\left(
\begin{array}{cccc}
\gamma _2\gamma _1 & -\gamma _2\gamma _1\beta _1 & -\gamma _2\beta
_2 & 0 \\ -\gamma _1\beta _1 & \gamma _1 & 0 & 0 \\ -\gamma
_2\gamma _1\beta _2 & \gamma _2\gamma _1\beta _2\beta _1 & \gamma
_2 & 0 \\ 0 & 0 & 0 & 1
\end{array}
\right) ,  \label{product}
\end{equation}

\bigskip

\noindent which is non-symmetric, as anticipated. Note that if one wants
to
speak about inertial systems, there are three of them here: the initial
system from which $\beta _1$ is defined, the second which is the result
of
applying the first boost and from which $\beta _2$ is measured and the
final
one obtained as a result of making the second boost. These systems are
all
taken with their spatial axis parallel to the previous one. These
considerations are not important in working out the computations, but
crucial when one wants to interpret them physically. So, we will write
equation $\left( \ref{product}\right) $ as the product of a boost $B_f$
and
a rotation $R$ \footnote{%
One could also opt for $B_f^{^{\prime }}R.$ The argument is the
same; note also that $RB_f=B_f^{^{\prime }}R$ implies
$B_f=R^TB_f^{^{\prime }}R.$} :
\begin{equation}
B_{(y)}B_{(x)}=RB_f,
\end{equation}
where
\begin{equation}
R=\left(
\begin{array}{cccc}
1 & 0 & 0 & 0 \\
0 & \cos \theta _W & \sin \theta _W & 0 \\
0 & -\sin \theta _W & \cos \theta _W & 0 \\
0 & 0 & 0 & 1
\end{array}
\right) .
\end{equation}
Therefore
\[
B_f=R^{-1}B_{(y)}B_{(x)}=\left(
\begin{array}{cccc}
1 & 0 & 0 & 0 \\
0 & \cos \theta _W & -\sin \theta _W & 0 \\
0 & \sin \theta _W & \cos \theta _W & 0 \\
0 & 0 & 0 & 1
\end{array}
\right) \left(
\begin{array}{cccc}
\gamma _2\gamma _1 & -\gamma _2\gamma _1\beta _1 & -\gamma _2\beta
_2 & 0 \\ -\gamma _1\beta _1 & \gamma _1 & 0 & 0 \\ -\gamma
_2\gamma _1\beta _2 & \gamma _2\gamma _1\beta _2\beta _1 & \gamma
_2 & 0 \\ 0 & 0 & 0 & 1
\end{array}
\right)
\]

\medskip\

\begin{equation}
\left(
\begin{array}{cccc}
\gamma _2\gamma _1 & -\gamma _2\gamma _1\beta _1 & -\gamma _2\beta
_2 & 0 \\ \left( -\gamma _1\beta _1\cos \theta _W+\gamma _2\gamma
_1\beta _2\sin \theta _W\right) & \left( \gamma _1\cos \theta
_W-\gamma _2\gamma  _1\beta _2\beta _1\sin \theta _W\right) &
-\gamma _2\sin \theta _W & 0 \\ \left( -\gamma _1\beta _1\sin
\theta _W-\gamma _2\gamma _1\beta _2\cos \theta _W\right) & \left(
\gamma _1\sin \theta _W+\gamma _2\gamma  _1\beta _2\beta _1\cos
\theta _W\right) & \gamma _2\cos \theta _W & 0 \\ 0 & 0 & 0 & 1
\end{array}
\right) .
\end{equation}

\bigskip

\noindent The angle $\theta _W$ can be obtained by demanding the
symmetry of the matrix $B_f$:
\begin{equation}
-\gamma _2\sin \theta _W=\gamma _1\sin \theta _W+\gamma _2\gamma
_1\beta _2\beta _1\cos \theta _W,
\end{equation}
i.e.
\begin{equation}
\tan \theta _W=-\frac{\gamma _2\gamma _1\beta _2\beta _1}{\gamma
_2+\gamma _1%
},  \label{wigner}
\end{equation}
or
\begin{equation}
\sin \theta _W=-\frac{\gamma _2\gamma _1\beta _2\beta _1}{\gamma
_2\gamma _1+1},\ \ \ \ \ \cos \theta _W=\frac{\gamma _2+\gamma
_1}{\gamma _2\gamma _1+1}.
\end{equation}
By replacing these values, one finds that the boost $B_f$ is

\begin{equation}
B_f=\left(
\begin{array}{cccc}
\gamma _2\gamma _1 & -\gamma _2\gamma _1\beta _1 & -\gamma _2\beta _2 &
0 \\
-\gamma _2\gamma _1\beta _1 & \left( 1+\frac{\gamma _2^2\gamma _1^2\beta
_1^2%
}{\gamma _2\gamma _1+1}\right) & \frac{\gamma _2^2\gamma _1\beta _2\beta
_1}{%
\gamma _2\gamma _1+1} & 0 \\
-\gamma _2\beta _2 & \frac{\gamma _2^2\gamma _1\beta _2\beta _1}{\gamma
_2\gamma _1+1} & \frac{\gamma _2(\gamma _2+\gamma _1)}{\gamma _2\gamma
_1+1}
& 0 \\
0 & 0 & 0 & 1
\end{array}
\right) ,  \label{boost2D}
\end{equation}

\medskip\

\noindent which is a boost along some direction in the $x-y$ plane. In
order
to find this direction, we will look for the direction in the $x-y$
plane
that is left invariant by the boost $B_f$; i.e., the direction that is
orthogonal to the direction of the boost. Since the vectors that are
orthogonal to the direction of the boost do not suffer changes (either
in
direction or magnitude), one can write $B_fw=w$ for such a four-vector
, or:
\begin{equation}
\left(
\begin{array}{cccc}
\gamma _2\gamma _1 & -\gamma _2\gamma _1\beta _1 & -\gamma _2\beta
_2 & 0 \\ -\gamma _2\gamma _1\beta _1 & \left( 1+\frac{\gamma
_2^2\gamma _1^2\beta
_1^2%
}{\gamma _2\gamma _1+1}\right) & \frac{\gamma _2^2\gamma _1\beta
_2\beta _1}{ \gamma _2\gamma _1+1} & 0 \\ -\gamma _2\beta _2 &
\frac{\gamma _2^2\gamma _1\beta _2\beta _1}{\gamma _2\gamma _1+1}
& \frac{\gamma _2(\gamma _2+\gamma _1)}{\gamma _2\gamma  _1+1} & 0
\\ 0 & 0 & 0 & 1
\end{array}
\right) \left(
\begin{array}{c}
0 \\
w^x \\
w^y \\
0
\end{array}
\right) =\left(
\begin{array}{c}
0 \\
w^x \\
w^y \\
0
\end{array}
\right) .
\end{equation}

\medskip\

As a consequence $\gamma _1\beta _1w^x+\beta _2w^y=0,$which can be
read by saying that the vector $w^x\hat x+w^y\hat y,$in the $x-y$
plane, is orthogonal to the vector $\gamma _1\beta _1\hat x+\beta
_2\hat y$ . Thus this last vector is in the direction of the boost
$B_f$ . In order to identify the velocity of the boost $B_f$, one
could consider the displacement four-vector between two events
that happen at the same place in the original coordinate system:
$\Delta =(\Delta \tau ,0,0,0)$, $\Delta \tau $ being the proper
time. Since $\Delta \rightarrow B_f\Delta $, then in the boosted
coordinate system the time interval between the events is $\gamma
_2\gamma _1\Delta \tau $. From the known relation between proper
time and coordinate time, one obtains the result that the gamma
factor (in other words, the velocity) of the boost $B_f$ is
$\gamma _f=\gamma _2\gamma _1$. Then $\beta _f^2=1-\gamma
_f^{-2}=1-\gamma _2^{-2}\gamma _1^{-2}=1-(1-\beta _2^2)(1-\beta
_1^2)=\beta _1^2+\gamma _1^{-2}\beta _2^2$. This result, together
with the direction of the boost, completes our understanding of
the transformation $B_f$.\footnote{ Alternatively, the velocity of
a boost $B(\vec \beta )$ can be straightforwardly read from the
first file of its matrix. Indeed, in order that the time
transformation adopts a form manifestly invariant under spatial
rotations ---$ct^{\prime }=\gamma (ct-\vec \beta \cdot \vec = r)$
---, the first file must be $(\gamma ,-\gamma \,\vec \beta ).$ }

In summary, the composition of a boost along the $x$ axis with velocity
$
\beta _1$ followed by a boost along the $y$ axis with velocity
$\beta_2$ is equivalent to a single boost with velocity $\vec
\beta _f=\beta _1\hat x +\gamma _1^{-1}\beta _2\hat y$ (the
relativistic composition of velocities), followed by a rotation in
the $x-y$ plane with angle $\theta _W=-\arctan \frac{\gamma
_2\gamma _1\beta _2\beta _1}{\gamma _2+\gamma _1}$ i.e.
\begin{eqnarray}
B_{(y)}(\beta _2)\ B_{(x)}(\beta _1)=R(\theta _W)\ B_f
\label{result}
\end{eqnarray}

where
\begin{equation}
\vec \beta _f=\beta _1\hat x+\gamma _1^{-1}\beta _2\hat y
\label{boostf}
\end{equation}

and as before

\begin{equation}
\tan \theta _W=-\frac{\gamma _2\gamma _1\beta _2\beta _1}{\gamma
_2+\gamma _1 }   \eqnum{8} \label{wigner}
\end{equation}

\bigskip\

As a preparation for the next Section, note that we can read
(\ref{result})
backward to note that any boost $B$ in the $x-y$ plane can be decomposed
into two mutually perpendicular boosts followed by a rotation:
\begin{equation}
B=R^{-1}\ B_{(y)}\ B_{(x)}  \label{resultbis}
\end{equation}

\bigskip

\subsection{Composition of arbitrary boosts}

Equipped with the previous understanding of the composition of two
perpendicular boosts, let us tackle the general case. A generic
composition of boosts can be seen as the composition of a boost
$B_{(a)}$ of velocity $ \vec \beta _a$, and a second boost $B$ of
velocity $\vec \beta =\vec \beta _{\Vert }+\vec \beta _{\perp }$,
where $\Vert $ and $\perp $ mean the parallel and perpendicular
directions with respect to the first boost $\vec \beta _a$. Since
the Wigner rotation is a geometric result (it only depends on the
velocities of the boosts and the angle between them), one is free
to choose the $x-y$ plane as the plane defined by both velocities,
the $x$ axis as the direction $\Vert $, and the $y$ axis as the
direction $\perp .$ Although a generic composition of boosts could
demand formidable algebraic manipulations, we will be able to get
the result by using only the results of the previous section. The
key to attaining our goal will be the decomposition Eq.
(\ref{resultbis}). In fact the main difficulty come from the fact
that the second boost has components $\hat x$ and $\hat y.$ Our
first step will consist in rewriting the second boost $B$ as a
composition of a boost along $\hat x$ and another boost along
$\hat y$. This was done formally at the end of the preceeding
section. We can thus use Eq. (\ref {resultbis}) to regard the
second boost $B(\vec \beta =\beta _{\Vert
}\hat x%
+\beta _{\perp }\hat y)$ as a product of a rotation and two mutually
perpendicular boosts, i.e.

\begin{equation}
B(\vec \beta )=R^{-1}\left( \phi \right) \ B_{(y)}(\beta _2\hat y)\
B_{(x)}(\beta _{\Vert }\hat x),
\end{equation}
where
\begin{equation}
\beta _2=\gamma _{\Vert }\beta _{\perp }  \label{rel1}
\end{equation}
in order that the relativistic composition of the velocities
$\beta _{\Vert } \hat x$ and $\beta _2\hat y$ gives back $\vec
\beta =\beta _{\Vert  }\hat x +\beta _{\perp }\hat y$ . Then
$\gamma _2=\gamma \gamma _{\Vert  }^{-1},$with $\gamma =\gamma
(\beta )$, and
\begin{equation}
\tan \phi =-\frac{\gamma _2\gamma _{\Vert }\beta _2\beta _{\Vert
}}{\gamma _2+\gamma _{\Vert }}=-\frac{\gamma \gamma _{\Vert }\beta
_{\perp  }\beta _{\Vert }}{\gamma \gamma _{\Vert }^{-1}+\gamma
_{\Vert }}.  \label{rel3}
\end{equation}

At first glance it would seem to the reader that we are going backward,
descomposing the boost instead of composing them. The advantage of doing
this will become clear in a few lines. We can now turn to the
composition of
$B(\vec \beta )$ and $B_{(a)}(\beta _a\hat x)$:

\begin{equation}
B(\vec \beta )\ B_{(a)}(\beta _a\hat x)=R^{-1}\left( \phi \right)
B_{(y)}(\beta _2\hat y)\ B_{(x)}(\beta _{\Vert }\hat x)\
B_{(a)}(\beta _a \hat x)=R^{-1}\left( \phi \right) B_{(y)}(\beta
_2\hat y)\ =
B_{(x)}(\beta _1%
\hat x),
\end{equation}
where

\begin{equation}
\beta _1=\frac{\beta _{\Vert }+\beta _a}{1+\beta _{\Vert }\beta _a}
\label{beta1}
\end{equation}
denotes the velocity corresponding to the composition of two
parallel boosts (then $\gamma _{1}=\gamma _{\Vert }\gamma
_a(1+\beta _{\Vert }\beta _a)$). Note that we combined the two
consecutive boost in the $\hat x$ direction using the well known
velocity addition formula. In this way one falls back to the
composition of the two remaining mutually perpendicular boosts. At
this point, let us recall our objective: we want to regard the
composition $%
B(\vec \beta )\ B_{(a)}(\beta _a\hat x)$ as the product of a
rotation $ R\left( \theta _W\right) $ in the $x-y$ plane and a
boost $B_f$. Then

\begin{equation}
R\left( \theta _W\right) B_f=B(\vec \beta )\ B_{(a)}(\beta _a\hat
x )=R^{-1}\left( \phi \right) \ B_{(y)}(\beta _2\hat y)\
B_{(x)}(\beta _1\hat x ),  \label{rbb}
\end{equation}
which means

\begin{equation}
R\left( \theta _W+\phi \right) B_f=B_{(y)}(\beta _2\hat y)\ =
B_{(x)}(\beta _1%
\hat x).  \label{sumofangle}
\end{equation}

The good new is that we have already solved this expression in the
previous
section! The matrix $B_f$ is that of (\ref{boost2D}) with the velocities
of (%
\ref{rel1}) and (\ref{beta1}). As shown there, $B_f$ is a boost whose
velocity $\vec \beta _f$ comes from the relativistic composition of the
velocities $\beta _1\hat x$ and $\beta _2\hat y$ :
\begin{equation}
\vec \beta _f=\beta _1\hat x+\gamma _1^{-1}\beta _2\hat
y=\frac{\beta _{\Vert }+\beta _a}{1+\beta _{\Vert }\beta _a}\hat
x+\frac{\gamma _a^{-1}\beta _{\perp }}{1+\beta _{\Vert }\beta
_a}\hat y,
\end{equation}
i.e. $\vec \beta _f$ is the relativistic composition of $\vec \beta _a$
and $%
\vec \beta $. The angle $\left( \theta _W+\phi \right) $ in Eq.(\ref
{sumofangle}) must satisfy the (\ref{wigner}):

\begin{equation}
\tan \left( \theta _W+\phi \right) =-\frac{\gamma _2\gamma _1\beta
_2\beta _1 }{\gamma _2+\gamma _1}=-\frac{\beta _{\perp }\left(
\beta _{\Vert  }+\beta _a\right) }{\gamma _{\Vert }^{-2}\gamma
_a^{-1}+\gamma ^{-1}\left( 1+\beta _{\Vert }\beta _a\right)
}\equiv \zeta .  \label{rel113}
\end{equation}

Since $\tan \left( \theta _W+\phi \right) =\left( \tan \theta
_W+\tan \phi \right) /(1-\tan \theta _W\tan \phi ),$ one concludes
that the Wigner rotation for the composition $B(\vec \beta =\beta
_{\Vert }\hat  x+\beta _{\perp }\hat y)\ B_{(a)}(\vec \beta
_a=\beta _a\hat x)$ is a rotation in the spatial plane defined by
the directions of both boosts, whose angle
$
\theta _W$ is given by

\begin{equation}
\tan \theta _W=\frac{\zeta -\tan \phi }{1+\zeta \tan \phi }.
\end{equation}
Recall that $\Vert $ and $\perp $ in these equations mean the
parallel and perpendicular directions with respect to the first
boost $\vec \beta _a$, in the spatial plane defined by both boosts
$\vec \beta _a$ and $\vec \beta $. The velocity $\vec \beta =\beta
_{\Vert }\hat x+\beta _{\perp }\hat y$ is measured by an observer
at rest in the system defined by the first boost
$%
\vec \beta _a$. Note that, $\zeta $ and $\phi $ are readily obtained
from
the data, namely $\beta _a,$ $\beta _{||}$ and $\beta _{\perp }$ via
Eqs. (%
\ref{rel113}) and (\ref{rel3}) .

\section{Aberration of light}

We will show an application of Wigner rotation in the context of
the aberration of light (i.e., the change of the propagation
direction of a light-ray produced by a boost). For simplicity we
shall work with two mutually perpendicular boosts. Let us choose
the $x$ axis to coincide with the propagation direction of the
light-ray. A first boost $B_{(x)}(\beta _1)$
leaves the propagation direction invariant, while a second boost $%
B_{(y)}(\beta _2)$ changes that direction according with the aberration
of
zenithal starlight law:
\begin{equation}
\delta _c=\arccos \gamma _2^{-1}  \label{abe}
\end{equation}
$\delta _c$ is the angle between the $x$ direction in the original
coordinate system (the light-ray) and the $x$ direction after the
composition. This is {\it not} the aberration angle due to a boost with
the
relativistically composed velocity $\vec \beta _f=\beta _1\hat
x+\gamma
_1^{-1}\beta _2\hat y$ . The Wigner rotation provides the difference
between
these two angles.

In fact, in Appendix 2 the aberration angle for a boost with velocity
$\vec
\beta _f=\beta _1\hat x+\gamma _1^{-1}\beta _2\hat y$ has been
computed; the
result is

\begin{equation}
\delta =\arccos \left[ \left( \beta _1^2+\beta _2^2\left( 1+\beta
_1\right)
\left( \gamma _2^{-1}\gamma _1^{-1}-\beta _1\right) \right) /\left(
\beta
_1^2+\gamma _1^{-2}\beta _2^2\right) \right]  \label{abe2}
\end{equation}
The difference between (\ref{abe}) and (\ref{abe2}) is due to the fact
that
the new $x$ direction in both process is not the same. So the boost
associated with the relativistically composed velocity $\vec \beta _f$
must
be completed with a rotation, in order to yield the aberration coming
from
the composition of boosts. The rotation angle $\delta -\delta _c$ is the
Wigner angle (\ref{wigner}). To make contact with our previous method,
what
we are saying is that in the first case:
\[
B_{(y)}\left( \beta _2\right) B_{(x)}\left( \beta _1\right) \left(
\begin{array}{c}
c \\
c \\
0 \\
0
\end{array}
\right) = \gamma _1\gamma _2\left( 1-\beta _1\right) \left(
\begin{array}{c}
c \\
c\cos \left( \delta _c\right) \\
c\sin \left( \delta _c\right) \\
0
\end{array}
\right) .
\]
while in the second case:
\begin{eqnarray*}
R\left( \theta _W\right) B_f\left( \beta _1\hat x+\gamma
_1^{-1}\beta _2\hat y\right) \left(
\begin{array}{c}
c \\
c \\
0 \\
0
\end{array}
\right) &=&R\left( \theta _W\right) \gamma _1\gamma _2\left( 1-\beta
_1\right) \left(
\begin{array}{c}
c \\
c\cos \left( \delta \right) \\
c\sin \left( \delta \right) \\
0
\end{array}
\right) \\
\ &=&\gamma _1\gamma _2\left( 1-\beta _1\right) \left(
\begin{array}{c}
c \\
c\cos \left( \delta -\theta _W\right) \\
c\sin \left( \delta -\theta _W\right) \\
0
\end{array}
\right) ,
\end{eqnarray*}
Since $B_{(y)}\left( \beta _2\right) B_{(x)}(\beta _1)=R\left( \theta
_W\right) B_f\left( \beta _1\hat x+\gamma _1^{-1}\beta _2\hat y\right)
,$
then $\delta -\theta _W=\delta _c$ as stated above. The multiplicative
factor $\gamma _1\gamma _2\left( 1-\beta _1\right) $ is the Doppler
shift.

\section{Conclusions}

Our argument for working out the Wigner rotation can then be given in a
nutshell as follows. First, a boost along the $x$ direction is
manifestly
symmetric. One can also understand this feature by noting that there are
two
null eigenvectors along the null cone (with eigenvalue equal to the
Doppler
shifts) and two trivial ones (along the $y$ and $z$ axis). Now, since a
generic boost is obtained by a rotation of the axis and $R^{-1}=R^T$
(that
is $R$ is orthogonal), the matrix representing a generic boost stays
symmetric (or, equivalently, it will preserve its four eigenvectors with
real eigenvalues). The symmetry allows us to easily compute the Wigner
angle
in the case of a composition of two perpendicular boosts. Now in the
generic
case, the problem can be cast in a form identical to the previous one,
after
carrying out a proper decomposition of the boosts into two mutually
perpendicular directions. Thus the answer is written without any
difficult
algebraic computing.

Physically not intuitive due to the lack of any Galilean analogue,
Wigner
rotation has been relegated to some corner of knowledge. Althought
Wigner
rotation is challenging both in terms of mathematical skill and physical
intuition, its computation is nonetheless within the reach of elementary
analysis and it is an instructive way to apprehend the subtlety inherent
to
the subject.

\medskip\

{\bf APPENDIX 1: Eigen-directions of a boost}

We will show the two null eigen-directions of a boost explicitly. Let
the
boost be in the $\hat x$ direction; dropping the two invariant spatial
directions $\hat y$ and $\hat z$ , and working just in the $t-x$ plane,
the
orthogonal transformation required is:

\begin{equation}
OB_x\left( \beta \right) O^T=\frac 1{\sqrt{2}}\left(
\begin{array}{cc}
1 & -1 \\
1 & 1
\end{array}
\right) \left(
\begin{array}{cc}
\gamma & -\gamma \beta \\
-\gamma \beta & \gamma
\end{array}
\right) \frac 1{\sqrt{2}}\left(
\begin{array}{cc}
1 & 1 \\
-1 & 1
\end{array}
\right) \left(
\begin{array}{cc}
\gamma \left( 1+\beta \right) & 0 \\
0 & \gamma \left( 1-\beta \right)
\end{array}
\right) .  \label{orto}
\end{equation}

The coordinate change is simply
\begin{eqnarray}
u &=&\frac 1{\sqrt{2}}\left( ct-x\right) , \\
v &=&\frac 1{\sqrt{2}}\left( ct+x\right) ,
\end{eqnarray}
which are the so-called null coordinates. The eigenvalues associated
with
the null directions are the relativistic Doppler shift factors (this is, of
course, not a surprising result). This change of coordinates is not a
Lorentz transformation, because it does not leave the Minkowski metric
invariant:
\begin{equation}
\frac 12\left(
\begin{array}{cc}
1 & -1 \\
1 & 1
\end{array}
\right) \left(
\begin{array}{cc}
-1 & 0 \\
0 & 1
\end{array}
\right) \left(
\begin{array}{cc}
1 & 1 \\
-1 & 1
\end{array}
\right) =\left(
\begin{array}{cc}
0 & -1 \\
-1 & 0
\end{array}
\right) .
\end{equation}
This is evident when we look at the transformation in a Minkowski
diagram:
this amount to a rigid rotation of $45^{\circ }$in the counter-clockwise
sense in space-time instead of the famous ''scissor-like'' picture of
the
Lorentz transformation. This can be traced to the fact that the proper
Lorentz group is isomorphic to $O(1,3)$ instead of $O(4).$The matrix $O$
in
Eq.(\ref{orto}) belongs to the group $O(4).$

\medskip\

{\bf APPENDIX 2: Computation of the aberration angle}

To begin with, we will recall the aberration angle due to a boost $%
B_{(x)}(\beta ).$ If the light-ray propagates in the direction $\hat n%
=\left( \cos \psi ,\sin \psi ,0\right) $, the transformed direction
$\hat n%
^{\prime }$ is obtained by applying the usual Lorentz
transformation to the velocity ${\bf \vec u}=c\hat n,$ which
transforms to ${\bf \vec u}^{\prime }=c\hat n^{\prime }$:
\begin{equation}
\hat n^{\prime }=\left( \frac{\cos \psi -\beta }{1-\beta \cos \psi
},\frac{ \sin \psi }{\gamma \left( 1-\beta \cos \psi \right)
},0\right) . \label{direction}
\end{equation}
The aberration angle is
\begin{equation}
\cos \delta =\hat n\cdot \hat n^{\prime }=\frac 1{1-\beta \cos \psi
}\left[
\cos \psi \left( \cos \psi -\beta \right) +\gamma ^{-1}\sin ^2\psi
\right] .
\label{scalar}
\end{equation}

In getting this result, the $x$ axis was chosen in the direction of the
boost because of practical reasons. But, of course, the aberration angle
depends only on the norm of $\vec \beta \,$ and the angle $\psi $
between $%
\vec \beta $ and the light-ray.

Let us now study the problem proposed in the body of the text. Let
there be a boost with velocity $\vec \beta _f=\beta _1\hat
x+\gamma
_1^{-1}\beta _2%
\hat y,$ and a light-ray traveling along the $x$ axis. Then, using the
substitutions
\[
\cos \psi =\frac{\beta _1}{\beta _f}=\frac{\beta _1}{\sqrt{\beta
_1^2+\gamma
_1^{-2}\beta _2^2}}\qquad ,\qquad \sin \psi =-\frac{\gamma
_1^{-1}\beta _2}{%
\beta _f}=-\frac{\gamma _1^{-1}\beta _2}{\sqrt{\beta _1^2+\gamma
_1^{-2}\beta _2^2}},
\]
in (\ref{scalar}) (the minus sign is due to the fact that the angle
$\psi $
is measured in the counter-clockwise sense from $\vec \beta _f$ to $\hat
n$%
), after some algebra one obtains:
\begin{equation}
\cos \delta =\frac{\beta _1^2+\beta _2^2\left( 1+\beta _1\right)
\left( \gamma _2^{-1}\gamma _1^{-1}-\beta _1\right) }{\beta
_1^2+\gamma _1^{-2}\beta _2^2},  \label{aberration}
\end{equation}
i.e. in the boosted system the angle between the light-ray (the $x$
direction in the original coordinate system) and the boost direction is
$%
\psi ^{\prime }=\psi +\delta .$

The result (\ref{aberration}) can be compared with that corresponding to
the
boost composition $B_{(y)}\left( \beta _2\right) B_{(x)}\left( \beta
_1\right) $. The first boost does not produce aberration, since it has
the
same direction as the light-ray. The second produces an aberration that
is a
particular case of (\ref{aberration}) with $\,\beta _1=0$ :
\begin{equation}
\cos \delta _c=\gamma _2^{-1}.  \label{aberration2}
\end{equation}
Of course the same result is recovered from (\ref{scalar}) by replacing
$%
\beta =\beta _2$ and $\psi =\pi /2.$

\smallskip\

{\bf ACKNOWLEDGMENTS}

The authors wish to thank Edgardo Garc\'\i a Alvarez and Daniel Sforza
for
reading the manuscript. This work was supported by Universidad de Buenos
Aires and Consejo Nacional de Investigaciones Cient\'\i ficas y
T\'ecnicas.


\begin{references}

\bibitem{jackson}  J.D. Jackson, {\it Classical Electrodynamics}, John Wiley
\& Sons Inc., N.Y., 1975.

\bibitem{wigner}  E.P. Wigner, Ann.Math. {\bf 40}, 149-204 (1939).

\bibitem{Goldstein}  H. Goldstein, {\it Classical Mechanics}, Addison-Wesley
Pub.
Co., 1980.

\bibitem{eisberg}  R.M. Eisberg, {\it Fundamentals of Modern Physics}, John
Wiley
\& Sons Inc.,1961.

\bibitem{taylor}  E.F. Taylor and J.A. Wheeler, {\it Spacetime Physics},
Freeman,
S.F., 1966.

\bibitem{barut}  A.O. Barut, {\it Electrodynamics and Classical Theory of
Fields and Particles}, Dover, 1980.

\bibitem{ajp1}  G.P. Fisher, Am.J.Phys {\bf 40}, 1772-1781 (1972).

\bibitem{ajp2}  A. Ben-Menahem, Am.J.Phys. {\bf 53}, 62-66 (1985).
\end{references}
\end{document}